\documentclass[a4paper,11pt]{article} 
\makeatletter
\let\@fnsymbol\@alph %@arabic  
\makeatother     
\usepackage[english]{babel}
\usepackage[a4paper]{geometry}
\usepackage{amsmath}
\usepackage{amsthm}
\usepackage{amssymb}
\usepackage{bbm}
\usepackage{color}
\usepackage{enumerate}
\usepackage{mathrsfs}

\usepackage{hyperref}         %%%% click to contents
\def\bR{\mathbb{R}}

\def\bN{\mathbb{N}}

\def\bZ{\mathbb{Z}}

\def\cM{\mathcal{M}}

\def\cO{\mathcal{O}}

\def\cF{\mathcal{F}}
\def\cG{\mathcal{G}}
\def\cL{\mathcal{L}}

\def\cN{\mathcal{N}}
\def\cE{\mathcal{E}}

\def\eps{\varepsilon}
\def\ph{\varphi}

\def\wt{\widetilde}

\def\indic{\hbox{\raise-2pt \hbox{\indbf 1}}}

\let\==\equiv

\let\io=\infty
\let\0=\noindent

\def\*{{\hfill\break\null\hfill\break}}

\def\ie{\hbox{\it i.e.\ }}

\def\tende#1{\,\vtop{\ialign{##\crcr\rightarrowfill\crcr
             \noalign{\kern-1pt\nointerlineskip}
             \hskip3.pt${\scriptstyle #1}$\hskip3.pt\crcr}}\,}
\def\otto{\,{\kern-1.truept\leftarrow\kern-5.truept\to\kern-1.truept}\,}

\newtheorem{theorem}{Theorem}[section]  % use thm for %Theorems to keep numbering consistent

\numberwithin{equation}{section}

\def\be{\begin{equation}}
\def\ee{\end{equation}}

                          \let\r=\rho
          \let\ph=\varphi   
        
        \let\L=\Lambda

\def\aa{\mathfrak{a}}
\def \blue#1 {\textcolor{blue}{#1}}
\def \red#1 {\textcolor{red}{#1}}

\definecolor{lightblue}{rgb}{0, 0.33, 0.71}

\title{Ground state energy of a Bose gas \\ in the Gross-Pitaevskii regime} 

\author{Giulia Basti$^{1,}$\footnote{Electronic mail: giulia.basti@gssi.it}\;, Serena Cenatiempo$^{1,}$\footnote{Electronic mail: serena.cenatiempo@gssi.it}, Alessandro Olgiati$^{2,}$\footnote{Electronic mail: alessandro.olgiati@math.uzh.ch}, \\ Giulio Pasqualetti$^{2,}$\footnote{Electronic mail: giulio.pasqualetti@math.uzh.ch}, and Benjamin Schlein$^{2,}$\footnote{Corresponding author: benjamin.schlein@math.uzh.ch} \\[0.2cm]
{\footnotesize $^{1}$Gran Sasso Science Institute, Viale Francesco Crispi 7, 67100 L'Aquila, Italy}\\
{\footnotesize $^{2}$Institute of Mathematics, University of Zurich, Winterthurerstrasse 190, 8057 Zurich.}
}

\begin{document} 

\maketitle

\begin{abstract}
We review some rigorous estimates for the ground state energy of dilute Bose gases. We start with Dyson's upper bound, which provides the correct leading order asymptotics for hard spheres. Afterwards, we discuss a rigorous version of Bogoliubov theory, which recently led to an estimate for the ground state energy in the Gross-Pitaevskii regime, valid up to second order, for particles interacting through integrable potentials. Finally, we explain how these ideas can be combined to establish a new upper bound, valid to second order, for the energy of hard spheres in the Gross-Pitaevskii limit. Here, we only sketch the main ideas, details will appear elsewhere. 
\end{abstract} 

\begin{center}
{\it This paper is dedicated to the memory of Freeman Dyson, whose work has \\ been 
inspiring the mathematical physics community for the last 70 years.}
\end{center} 

\section{Introduction}

In 1957, in a short but very influential paper \cite{Dy}, Dyson proved the first rigorous estimate for the ground state energy of an interacting Bose gas at low density. He considered  a gas of hard spheres, but his approach can be easily extended to more general  repulsive, short-range interactions. To leading order, he obtained the correct upper bound for the ground state energy and a lower bound, off by a factor around fourteen. 

Forty years later, in 1997, Lieb-Yngvason \cite{LY} managed to show a lower bound for the ground state energy, matching Dyson's upper bound. They considered a system of $N$ bosons, moving in the box $\Lambda_L = [0; L]^3$ (with periodic boundary conditions) and interacting through a repulsive (ie. non-negative),  two body radial potential $V : \bR^3 \to \bR$, with integrable decay at infinity. They proved that the ground state energy per unit particle 
 at fixed density $\rho=N/L^3$, in the limit $N,L \to \io$, is such that 
\begin{equation}\label{eq:LY} e (\rho) = 4 \pi \frak{a} \rho \, \big( 1 + o(1) \big) \end{equation} 
in the dilute limit $\rho \frak{a}^3 \to 0$. Here $\frak{a}$ denotes the scattering length of the potential $V$ and is defined through the solution of the zero-energy scattering equation 
\begin{equation}\label{eq:0-en} \Big[ -\Delta + \frac{1}{2} V \Big] f = 0 \nonumber
\end{equation} 
with the boundary condition $f (x) \to 1$, as $|x| \to \infty$, so that $f(x) \simeq 1 - \frak{a}/|x|$, asymptotically for large $|x|$. For the hard-sphere interaction, $\frak{a}$ coincides with the radius of the spheres. 

The next order corrections to (\ref{eq:LY}) have been predicted in the physics literature by Lee-Huang-Yang \cite{LHY}, who derived the expression
\begin{equation}\label{eq:LHY}  e(\rho) = 4\pi\frak{a} \rho \,  \Big[ 1 + \frac{128}{15 \sqrt{\pi}} (\rho \frak{a}^3)^{1/2} + o \big( (\rho \frak{a}^3)^{1/2} \big) \Big] \nonumber
\end{equation} 
again for the dilute limit $\rho \frak{a}^3 \ll 1$. It is interesting to observe how (\ref{eq:LY}) and also (\ref{eq:LHY}) only depend on the scattering length $\frak{a}$, and not on further details of the interaction. A rigorous lower bound matching the Lee-Huang-Yang formula has been recently derived by Fournais-Solovej, first for integrable potentials in \cite{FS1} and then also for hard-sphere interactions in \cite{FS2}. A rigorous upper bound for the ground state energy per particle matching (\ref{eq:LHY}), on the other hand, has been first derived by Yau-Yin in \cite{YY} for smooth interaction potentials  (a quasi-free trial state, which only recovered the correct asymptotics in the limit of weak potential, was previously proposed in  \cite{ESY}). More recently, a simpler trial state has been proposed in \cite{BCS}, for interaction potentials $V \in L^3 (\bR^3)$. Still missing is an upper bound matching (\ref{eq:LHY}), for non-integrable interactions, like hard-sphere potentials (ironically, the upper bound is still missing exactly for the hard-sphere interaction that was first considered by Dyson, who probably regarded it as the simplest case).  

In experiments, Bose gases are often very dilute. This leads us to consider also scaling regimes in which the density is coupled with the number of particles $N$, and tends to $0$, as $N \to \infty$. An important example is the Gross-Pitaevskii regime, where the density is  $1/N^{2}$. Rescaling lengths, the Gross-Pitaevskii regime corresponds to systems of $N$ particles moving in the fixed box $\Lambda = [0;1]^3$ (with periodic boundary conditions) and interacting through a repulsive (ie. non-negative) rescaled potential $N^2 V (N \cdot)$, whose scattering length is given by $\frak{a}/N$, with $\frak{a}$ denoting the scattering length of $V$. Translating (\ref{eq:LY}) to the Gross-Pitaevskii limit, it implies that the ground state energy in this limit is given by 
\begin{equation}\label{eq:lead-GP} E_N^\text{GP} = 4\pi \frak{a} N  + o (N) 
\end{equation} 
as $N \to \infty$. 

In the last years, making use of a rigorous version of Bogoliubov theory \cite{B}, it was possible, at least for integrable potentials, to go beyond the leading order estimate (\ref{eq:lead-GP}) and to resolve the ground state energy and the low-energy excitation spectrum up to errors vanishing in the limit of large $N$. For particles interacting through a repulsive, radial and compactly supported interaction $V \in L^3 (\bR^3)$, the ground state energy in the Gross-Pitaevskii regime was shown in \cite{BBCS4} to satisfy 
\begin{equation}\label{eq:ENGP} \begin{split} E^{\text{GP}}_{N} = \; &4\pi \frak{a} (N-1) + e_\Lambda \frak{a}^2 \\ & - \frac{1}{2}\sum_{p\in\Lambda^*_+} \left[ p^2+8\pi \frak{a}  - \sqrt{|p|^4 + 16 \pi \frak{a}  p^2} - \frac{(8\pi \frak{a})^2}{2p^2}\right] + \cO (N^{-1/4}) \,  
\end{split}
\end{equation}   
where 
\begin{equation}\label{eq:eLambda0}
e_\Lambda = 2 - \lim_{M \to \infty} \sum_{\substack{p \in \bZ^3 \backslash \{ 0 \} : \\ |p_1|, |p_2|, |p_3| \leq M}} \frac{\cos (|p|)}{p^2} \end{equation}
and, in particular, the limit can be proven to exist. The second line of (\ref{eq:ENGP}) corresponds, in the thermodynamic limit, to the second term in (\ref{eq:LHY}) (in this regime, the sum in (\ref{eq:ENGP}) can be replaced by an integral, leading to (\ref{eq:LHY}); see \cite[Eqs. (A.26)-(A.29)]{LSSY}). The correction $e_\Lambda \frak{a}^2$, on the other hand, is a finite volume effect, arising because $\frak{a}$ is defined through the scattering equation (\ref{eq:0-en}) on the whole space $\bR^3$, rather than on the box $\Lambda$, and it has no analogue in (\ref{eq:LHY}). 

Additionally, after subtraction of the ground state energy, the excitation spectrum below a threshold $\zeta > 0$ was proven in \cite{BBCS4} to consist of eigenvalues having the form 
 \begin{equation}
    \begin{split}\label{eq:excGP}
    \sum_{p\in 2\pi \bZ^3 \backslash \{ 0 \}} n_p \sqrt{|p|^4+ 16 \pi \frak{a}  p^2}+ \cO (N^{-1/4} \zeta^3) \,, 
    \end{split}
    \end{equation}
with $n_p \in \bN$ for all momenta $p \in 2\pi \bZ^3 \backslash \{ 0 \}$. In other words, excited eigenvalue are determined, in good approximation, by the sum of the energies of  quantized excitations, labelled by their momentum and characterized by the dispersion law $\eps (p) = \sqrt{|p|^4 + 16 \pi \frak{a} p^2}$. Again, we observe that (\ref{eq:ENGP}) and (\ref{eq:excGP}) only depend on the interaction potential through its scattering length $\frak{a}$. 

The results that we discussed in the last two paragraph, concerning particles trapped in the unit torus $\L$ in the Gross-Pitaevskii limit, can also be extended to Bose gases on $\bR^3$, confined by external trapping potentials $V_\text{ext}$. In this case, Lieb-Seiringer-Yngvason proved in \cite{LSY} that the ground state energy per particle is given, to leading order, by 
\[ \lim_{N\to \infty} \frac{E_N^\text{GP} (V_\text{ext})}{N} = \min_{\ph \in L^2 (\bR^3) : \| \ph \| =1} \cE_\text{GP} (\ph)\,, \]
with the Gross-Pitaevskii energy functional 
\[ \cE_\text{GP} (\ph) = \int \Big[ |\nabla \ph (x)|^2 + V_\text{ext} (x) |\ph (x) |^2 + 4\pi \frak{a} |\ph (x)|^4 \Big] dx\,. \]
More precise bounds, analogous to (\ref{eq:ENGP}) and (\ref{eq:excGP}), have been recently established, for integrable potentials, in \cite{NT,BSS2}. 

The estimates (\ref{eq:ENGP}), (\ref{eq:excGP}) have been proven in \cite{BBCS4} for $V \in L^3 (\bR^3)$. As indicated in \cite{NT}, it would be relatively easy to extend them to $V \in L^1 (\bR^3)$. On the other hand, extension to hard-spheres (or to more general non-integrable potentials with a hard-core) requires substantial new ideas. The goal of these notes is to present an upper bound for the ground state energy in the Gross-Pitaevskii regime, for particles interacting through a hard-sphere potential, consistent with (\ref{eq:ENGP}); a complete proof of this estimate will appear in \cite{BCOPS}, here 
we will only sketch some of the main ideas. 

Before stating and discussing our new estimate in Section \ref{sec:new}, we are going to review some tools that play a crucial role in its proof. In Section \ref{sec:dyson}, we will briefly recall Dyson's upper bound for the leading order contribution to the ground state energy of a dilute gas of hard spheres. In Section \ref{sec:bog}, on the other hand, we will review some of the main steps of the rigorous version of Bogoliubov theory for integrable potentials that led to (\ref{eq:ENGP}), (\ref{eq:excGP}).

\section{Dyson's upper bound for a dilute Bose gas} 
\label{sec:dyson}

In this section, we review Dyson's upper bound for the ground state energy of hard-spheres, in the dilute limit. Since the argument plays an important role in the derivation of our new estimate for the Gross-Pitaevskii regime, we are going to sketch a slightly different proof, compared with Dyson's original work (in fact, our trial state is closer to the one proposed by Jastrow in \cite{J}).  

We consider a gas of $N$ hard spheres  in the three dimensional torus $\Lambda_L = [0;L]^{3}$. We are looking for an upper bound to the ground state energy 
\begin{equation}\label{eq:ENL} E_{N,L} = \inf \;  \frac{\langle \Psi, \sum_{j=1}^N -\Delta_{x_j} \Psi \rangle}{\| \Psi \|^2} \end{equation}
with the infimum taken over all $\Psi \in L^2 (\L^N_L)$, symmetric with respect to permutations of the $N$ particles and satisfying the hard-sphere condition $\Psi (x_1, \dots , x_N) = 0$, if there exist $i , j \in \{ 1, \dots , N \}$, $i \not = j$, with $|x_i - x_j| < \frak{a}$.  Here (and in the following) we indicate with $|x_i-x_j|$  the distance between $x_i$ and $x_j$ on the torus. We are interested in the thermodynamic limit, where $N, L \to \infty$ with fixed $\rho = N/ L^3$, and we focus on the low density regime, where $\rho \frak{a}^3 \ll 1$. 

To get an upper bound for (\ref{eq:ENL}), we have to evaluate the energy of an appropriate trial state. To this end, we modify the non interacting ground state $\Psi_{N,L}^{\frak{a} = 0} (x_1, \dots , x_N) \equiv 1$, by adding correlations among particles. Since correlations are produced mainly by two-body scattering events, it seems natural to consider trial states having the form
\begin{equation}\label{eq:psiNL}  \Psi_{N,L} (x_1, \dots , x_N) = \prod_{i<j}^N f_\ell (x_i - x_j) \end{equation}
where $\frak{a} \ll \ell \ll L$ is a parameter that will be fixed later and $f_\ell$ is meant to describe two-body correlations up to distance $\ell$. Product trial functions like (\ref{eq:psiNL}) have been first considered by Jastrow in \cite{J} (Dyson, on the other hand, worked in \cite{Dy} with a non-symmetric trial state, describing only nearest neighbour correlations). More precisely, in (\ref{eq:psiNL}) we choose $f_\ell$ as the ground state solution of the Neumann problem 
\begin{equation}\label{eq:neu1} -\Delta f_\ell = \lambda_\ell f_\ell \end{equation} 
on the ball $|x| \leq \ell$, with the hard-sphere condition $f_\ell (x) = 0$ for $|x| < \frak{a}$ and with the normalization $f_\ell (x) = 1$ for $|x| = \ell$. 
We can then extend $f_\ell$ to a function over the torus $\Lambda_L$, setting $f_\ell (x) = 1$ for $|x| > \ell$  \footnote{We cutoff correlations by choosing $f_\ell$ as the solution of the Neumann problem on the ball $|x| \leq \ell$. Compared with the simpler truncation of the zero-energy scattering equation used in \cite{Dy}, our procedure has the advantage that the computation of the many-body kinetic energy (\ref{eq:Psi-kin}) produces a nice two-body potential (in \cite{Dy}, the correct energy emerges as a boundary term, corresponding to a potential supported on $|x| = \ell$). This will play an important role in Section  \ref{sec:new}, where we will discuss how to modify (\ref{eq:psiNL}) to obtain better estimates.}. 
With standard tools in analysis (see, for example, \cite[Lemma A.1]{ESY-hierarchy}), we can study the eigenvalue problem (\ref{eq:neu1}) and prove that  
\begin{equation}\label{eq:lambda} \lambda_\ell = \frac{3\frak{a}}{\ell^3} \Big[ 1 + \cO (\frak{a}/\ell)\,. \Big] \end{equation} 
Moreover, denoting by $\chi_\ell$ the characteristic function of the ball $|x| \leq \ell$, we find 
\begin{equation}\label{eq:uell} |\nabla f_\ell (x)| \leq C \frak{a} \frac{\chi_\ell (x)}{|x|^2} , \qquad  0 \leq 1 - f_\ell^2 (x) \leq C \frak{a} \frac{\chi_\ell (x)}{|x|}\,. \end{equation} 

To compute the energy of the trial state (\ref{eq:psiNL}), we observe that 
\begin{equation}\label{eq:dys-comp} \begin{split} 
\frac{-\Delta_{x_j} \Psi_N (x_1, \dots , x_N) }{\prod_{i<j}^N f_\ell (x_i - x_j) } =  \sum_{i \not = j}^N \frac{-\Delta f_\ell (x_j - x_i)}{f_\ell (x_j - x_i)}  - \sum^N_{i,m,j} \frac{\nabla f_\ell (x_j - x_i)}{f_\ell (x_j  -x_i)} \cdot \frac{\nabla f_\ell (x_j - x_m)}{f_\ell (x_j - x_m)}  \,,
\end{split} \end{equation} 
where the sum in the last term runs over $i,j, m \in \{1, \dots , N \}$ all different. From (\ref{eq:neu1}), we obtain, with the notation ${\bf x} = (x_1, \dots, x_N) \in \L^N_L$, 
\begin{equation}\label{eq:Psi-kin}  \begin{split} \langle \Psi, \sum_{j=1}^N -\Delta_{x_j} \Psi \rangle = \; &2\lambda_\ell \sum_{i<j}^N \int \chi_\ell (x_i - x_j) \prod_{i<j}^N f_\ell^2 (x_i - x_i) d{\bf x} \\ &- \sum_{i,j,m} \int \frac{\nabla f_\ell (x_j - x_i)}{f_\ell (x_j - x_i)} \cdot \frac{\nabla f_\ell (x_j - x_m)}{f_\ell (x_j - x_m)}  \prod_{i<j}^N f^2_\ell (x_i - x_j) d{\bf x}\,. \end{split} \end{equation} 
Thus, using the permutation symmetry, we find 
\[ \begin{split} \frac{\langle \Psi, \sum_{j=1}^N -\Delta_{x_j} \Psi \rangle}{\| \Psi \|^2} = \; &N(N-1)\lambda_\ell \frac{ \int \chi_\ell (x_1 - x_2) \prod_{i<j}^N f_\ell^2 (x_i - x_j) d{\bf x}}{ \int  \prod_{i<j}^N f_\ell^2 (x_i - x_j) d{\bf x}} \\ &- N (N-1)(N-2) \frac{\int \frac{\nabla f_\ell (x_1 - x_3)}{f_\ell (x_1 - x_3)} \cdot \frac{\nabla f_\ell (x_2 - x_3)}{f_\ell (x_2 - x_3)} \prod_{i<j}^N f_\ell^2 (x_i - x_j) d{\bf x}}{  \int \prod_{i<j}^N f_\ell^2 (x_i - x_j) d{\bf x}} \\ = \; &\text{A} + \text{B}\,. \end{split} \]
Defining $u_\ell (x) = 1- f^2_\ell (x)$ and estimating 
\[ 1 - \sum_{j=2}^N u_\ell (x_1 - x_j)  \leq \prod_{j=2}^N f^2_\ell (x_1 - x_j) \leq 1 \]
in the numerator and in the denominator of $\text{A}$, we can decouple the variable $x_1$; with (\ref{eq:lambda}), (\ref{eq:uell}), we obtain  
\[ \text{A} \leq N^2 \lambda_\ell \frac{\int \chi_\ell (x) dx}{L^3 - N \int u_\ell (x) dx} \leq 4\pi \frak{a} \rho N \cdot \Big[ 1 + C \frac{\frak{a}}{\ell} + C \rho \frak{a} \ell^2 \Big] \]
under the assumption that $\frak{a}/ \ell , \rho \frak{a} \ell^2 \ll 1$. Similarly, using again (\ref{eq:uell}), we can bound (this estimate could be improved using that, by symmetry, the integral of $\nabla f_\ell (x_1- x_2) \cdot \nabla f_\ell (x_2- x_3)$ over $x_1, x_2, x_3$ vanishes) 
\[ \text{B} \leq C N^3 \frac{ \Big[ \int |\nabla f_{N,\ell} (x)| dx \Big]^2}{L^6 - C N L^3 \int u_\ell (x) dx} \leq \frac{C N^3 \frak{a}^2 \ell^2}{L^6 (1-C \r \aa \ell^2)} \leq C N \rho \frak{a} \big[ \rho \frak{a} \ell^2 \big]\,. \]
Thus, we conclude that 
\[ E_{N,L} \leq 4 \pi \frak{a} \rho N \cdot \Big[ 1 + C \frac{\frak{a}}{\ell} + C \rho \frak{a} \ell^2 \Big]\,. \]
Choosing $\ell = \rho^{-1/3}$, we obtain the upper bound 
\begin{equation*}\label{eq:up-fin} E_{N,L} \leq 4 \pi \frak{a} \rho N \cdot \Big[ 1 + C \frak{a} \rho^{1/3} \Big]\,, \end{equation*} 
which captures the correct leading order of the ground state energy $E_{N,L}$, in the dilute limit.

\section{Bogoliubov theory for trapped Bose gases} 
\label{sec:bog}

In this section, we consider a gas of $N$ bosons moving  in the unit torus $\Lambda = [0;1]^3$  and interacting through a repulsive (ie. non-negative) potential with scattering length of the order $1/N$ (Gross-Pitaevskii regime).
The Hamilton operator takes the form  
\begin{equation}\label{eq:HNGP} H^\text{GP}_N = \sum_{j=1}^N -\Delta_{x_j} + \sum_{i<j}^N N^2 V (N (x_i - x_j)) \end{equation}
and acts on the Hilbert space $L^2_s (\Lambda^N)$, the subspace of $L^2 (\L^N)$ consisting of functions that are symmetric with respect to permutations of the $N$ particles. 
Here, we assume $V: \bR^3 \to \bR$ to be a repulsive, radial and compactly supported potential. 

In fact, it is convenient to embed (\ref{eq:HNGP}) in a family of Hamilton operators for trapped Bose gases, all acting on $L^2_s (\L^N)$, given by 
\begin{equation}\label{eq:HNbeta} H^\beta_N = \sum_{j=1}^N -\Delta_{x_j} + \frac{1}{N} \sum_{i<j}^N N^{3\beta} V (N^\beta (x_i - x_j)) \end{equation}
and depending on a parameter $\beta \in [0;1]$. For $\beta = 1$, we recover the Gross-Pitaevskii Hamiltonian (\ref{eq:HNGP}). For $\beta = 0$, on the other hand, (\ref{eq:HNbeta}) describes a mean-field scaling, where particles interact weakly through a potential varying on the same length scale $L=1$ characterizing the box $\Lambda$. Mathematically, the mean-field scaling is more approachable, but from the point of view of physics, the Gross-Pitaevskii regime is of course more relevant. 

Extending the bound (\ref{eq:lead-GP}) to all $\beta \in [0;1]$, we find that the ground state energy of (\ref{eq:HNbeta}) is given, to leading order, by 
\[ \lim_{N \to \infty} \frac{E_N^\beta}{N} = 4 \pi \lim_{N \to \infty} \frak{a}^\beta_N  \] 
with $\frak{a}^\beta_N$ denoting $N$ times the scattering length of the rescaled interaction  potential $N^{3\beta-1} V (N^\beta \cdot)$. For $\beta = 1$, simple scaling of the zero-energy scattering equation (\ref{eq:0-en}) implies that $\frak{a}_N^{\beta =1} = \frak{a}$; for $\beta < 1$, on the other hand, $\frak{a}_N^\beta$ converges towards its first Born approximation $\widehat{V} (0)/ 8\pi$ (the Born series for $\frak{a}_N^\beta$ is given below, in (\ref{bN})). We conclude that 
\begin{equation}   \label{eq:LSY0} \lim_{N \to \infty} \frac{E_N^\beta}{N} = \left\{ \begin{array}{ll} \widehat{V} (0)/2 &\text{if } \beta \in [0;1) \\ 4 \pi \frak{a} &\text{if } \beta = 1 \end{array} \right. \,. \end{equation}  

In the last decade, there has been progress in the mathematical understanding of the properties of trapped Bose gases described by (\ref{eq:HNbeta}), based on rigorous versions of Bogoliubov theory, that made it possible to go beyond the leading order estimate (\ref{eq:LSY0}), at least for certain classes of integrable interactions. In the rest of this section, we are going to describe some of the main new ideas and results in this area. We focus here on integrable interaction potentials, excluding hard-spheres (we will come back to this point at the very end of this section and in the next). 

The first important observation in Bogoliubov theory is that, close to the ground state energy, trapped Bose gases exhibit complete Bose-Einstein condensation. All particles, up to a fraction vanishing as $N \to \infty$, are described by the same orbital $\ph_0$ defined by $\ph_0 (x) = 1$, for all $x \in \L$. In the mean-field limit and assuming the interaction potential to be bounded and positive definite, Bose-Einstein condensation follows from 
\[ \begin{split}  0 &\leq \int dx dy  \, V(x-y) \left[ \sum_{j=1}^N \delta (x -x_j) - N \right] \left[ \sum_{i=1}^N \delta(y-x_i) - N \right] \\ &= \sum_{i,j=1}^N V(x_i - x_j) - N^2 \widehat{V} (0) = 2 \sum_{i<j}^N V(x_i -x_j)  + N V(0) - N^2 \widehat{V} (0) \end{split} \]
which immediately gives
\[ H_N^{\beta = 0} \geq \frac{N}{2} \widehat{V} (0) + \sum_{j=1}^N -\Delta_{x_j} - C \,.\]
This operator inequality implies that, for every approximate ground state $\psi_N$, ie. for every $\psi_N \in L^2_s (\L^N)$ satisfying 
\[ \big\langle \psi_N, \big[ H_N^{\beta=0} - N \widehat{V}(0)/2 \big] \psi_N \big\rangle \leq C , \]
the expected number of orthogonal excitations of the condensate is bounded by  
\begin{equation}\label{eq:BECstr} \langle \psi_N, \big[ \frak{q}_1 + \dots + \frak{q}_N \big] \psi_N \rangle \leq C\,, \end{equation} 
where we introduced the notation $\frak{q} = 1- |\ph_0 \rangle \langle \ph_0 |$ for the projection on the orthogonal complement of the condensate wave function $\ph$ and where we used the gap in the Laplace operator to estimate $-\Delta \geq C \frak{q}$. Remark that (\ref{eq:BECstr}) implies condensation in a strong sense, proving that the number of excitations remains bounded, as $N \to \infty$.  

Verifying the existence of Bose-Einstein condensation in the Gross-Pitaevskii regime (more generally, for $\beta > 1/3$, ie. if the range of the potential is much smaller than the typical distance between particles) is much more challenging. The first proof was given by Lieb-Seiringer. In \cite{LS,LS2}, they showed that every approximate ground state wave function of (\ref{eq:HNGP}) is such that 
\[ \frac{1}{N} \langle \psi_N, \big[ \frak{q}_1 + \dots + \frak{q}_N \big] \psi_N \rangle \to 0 \]
as $N \to \infty$. A similar result was later obtained in \cite{NRS}, with different tools. More recently, the stronger estimate (\ref{eq:BECstr}), giving optimal bounds on the number of excitations, was proven to hold in the Gross-Pitaevskii regime and, in fact, for all $\beta \in (0;1]$, in \cite{BBCS1,BBCS4}. Different proofs and extensions to the case of particles trapped by external potentials have been obtained in \cite{NNRT,H,BSS}.

After establishing Bose-Einstein condensation, the next step in Bogoliubov theory consists in factoring out the condensate and in focusing instead on its orthogonal excitations. To this end, we proceed as in \cite{LNSS} and we observe that every $\psi_N \in L^2_s (\L^N)$ can be written as 
\[ \psi_N = \alpha_0 \ph_0^{\otimes N} + \alpha_1 \otimes_s \ph^{\otimes (N-1)} + \dots + \alpha_{N-1} \otimes_s \ph_0 + \alpha_N \]
with uniquely determined $\alpha_j \in L^2_\perp (\L)^{\otimes_s j}$, for $j=0,\dots, N$ (here $\otimes_s$ denotes the symmetric tensor product). Defining $U_N \psi_N = \{ \alpha_0, \dots, \alpha_N \}$, we map the Hilbert space $L^2_s (\L^N)$ into the truncated Fock space 
\[ \cF^{\leq N}_+ = \bigoplus_{n=0}^N L^2_\perp (\L)^{\otimes_s n} \]
constructed over the orthogonal complement $L^2_\perp (\L)$ of the condensate wave function $\ph_0$. 
Through $U_N$, we can define the excitation Hamiltonian $\cL_N^\beta = U_N H_N^\beta U_N^*$, acting on the excitation Hilbert space $\cF^{\leq N}_+$. 

To compute $\cL_N^\beta$, it is convenient to rewrite (\ref{eq:HNbeta}) in momentum space, using the formalism of second quantization with the creation and annihilation operators $a_p^*, a_p$, defined for every $p \in \L^* = 2\pi \bZ^3$, satisfying canonical commutation relations
\[ \big[ a_p , a_q^* \big]= \delta_{p,q} , \qquad \big[ a_p, a_q \big] = \big[ a_p^* , a_q^* \big] = 0\,. \]
We find 
\begin{equation*} \label{eq;HNkappa} 
H_N^\beta = \sum_{p \in \L^*_+} p^2 a_p^* a_p + \frac{1}{2N} \sum_{p,q,r \in \L^*} \widehat{V} (r/N^\beta) a_{p+r}^* a_q^* a_{q+r} a_p \,.
\end{equation*}
Denoting by $\cN_+$ the number of particles operator on $\cF^{\leq N}_+$ ($\cN_+$ measures the number of excitations of the condensate), we have, from \cite{LNSS}, the rules  
\begin{equation}\label{eq:rules} \begin{split} U_N a_0^* a_0 U^*_N &= N - \cN_+  \\
U_N a_0^* a_p U_N^* &= \sqrt{N- \cN_+} a_p =: \sqrt{N} b_p \\
U_N a_p^* a_0 U_N^* &= a_p^* \sqrt{N- \cN_+} =: \sqrt{N} b_p^* \\
U_N a_p^* a_q U_N &= a_p^* a_q  \end{split} \end{equation} 
for all $p,q \in \L^*_+ = 2\pi \bZ^3 \backslash \{ 0 \}$. Thus, we obtain 
\[ \cL_{N,\beta} = U_N H_N^\beta U_N^* = \cL^{(0)}_{N,\beta}  + \cL^{(2)}_{N,\beta} + \cL^{(3)}_{N,\beta} + \cL^{(4)}_{N,\beta}\,, \]
where we defined    
\begin{equation}\label{eq:cLN} \begin{split}  \cL_{N,\beta}^{(0)} = \; &\frac{N-1}{2N} \widehat{V} (0) (N- \cN_+) + \frac{\widehat{V} (0)}{2N} \cN_+ (N - \cN_+) \\
\cL_{N,\beta}^{(2)} =\; &  \sum_{p \in \Lambda^*_+} p^2 a_p^* a_p + \sum_{p \in \Lambda_+^*} \widehat{V} (p/N^\beta) a_p^* \frac{N-\cN_+-1}{N} a_p + \frac{1}{2} \sum_{p \in \Lambda^*_+} \widehat{V} (p/N^\beta) (b_p^* b_{-p}^* + \text{h.c.} ) \\
\cL_{N,\beta}^{(3)} = \; &\frac{1}{\sqrt{N}} \sum_{p,q \in \Lambda_+^* , p+q \not = 0} \widehat{V} (p/N^\beta) \left[ b_{p+q}^* a_{-p}^* a_q + a_q^* a_{-p} b_{p+q} \right]  \\
\cL_{N,\beta}^{(4)} = \; & \frac{1}{2N} \sum_{p,q \in \Lambda^*_+ , r \not = p , -q} \widehat{V} (r/N^\beta) a_{p+r}^* a_q^* a_p a_{q+r}\,. \end{split} \end{equation} 
In (\ref{eq:rules}), we introduced operators $b_p^*, b_p$ creating and, respectively, annihilating an excitation with momentum $p$ (removing and, respectively, adding a particle to the condensate). On states with only few excitations, ie. with $\cN_+ \ll N$, we have $b_p^* \simeq a_p^*$ and $b_p \simeq a_p$.  

The cubic and the quartic terms in (\ref{eq:cLN}) look small, in the limit $N \to \infty$. In the mean-field regime, one can prove that, on states with few excitations, their contribution is indeed negligible. In other words, for $\beta = 0$, we can approximate 
\[ \begin{split} \cL_{N,0} &\simeq  \cL_{N,0}^{(0)} + \cL^{(2)}_{N,0}  \\ &= \frac{(N-1)}{2} \widehat{V} (0) + \sum_{p \in \L^*_+} \big[ p^2 + \widehat{V} (p) \big] a_p^* a_p + \frac{1}{2} \sum_{p \in \Lambda^*_+} \widehat{V} (p) (b_p^* b_{-p}^* + \text{h.c.} )\,. \end{split}  \]
To diagonalize the resulting quadratic operator,  we conjugate it with a (generalized) Bogoliubov transformation having the form 
\begin{equation} \label{eq:Ttau} T_\tau = \exp \Big[ \frac{1}{2} \sum_{p \in \L^*_+} \tau_p (b_p^* b_{-p}^* - b_p b_{-p} ) \Big]\,.\end{equation}
On states with few excitations, we expect (\ref{eq:Ttau}) to act almost as a standard Bogoliubov transformation (with $b^\sharp_{\pm p}$ replaced by $a^\sharp_{\pm p}$), ie. 
\[ T_\tau^* a_p T_\tau  \simeq \cosh (\tau_p) \, a_p + \sinh (\tau_p) \, a_{-p}^*\,. \]
Choosing $\tau \in \ell^2 (\L^*_+)$ so that $\tanh (2\tau_p) = - \widehat{V} (p) / (p^2 + \widehat{V} (p))$, we obtain therefore 
\[ \begin{split} T_\tau^* \cL_{N,0} T_\tau \simeq \; &\frac{(N-1)}{2} \widehat{V} (0) - \frac{1}{2} \sum_{p \in \L^*_+} \Big[ p^2 + \widehat{V} (p) -\sqrt{|p|^4 + 2 p^2 \widehat{V} (p)} \Big] \\ &+ \sum_{p \in \L^*_+} \sqrt{|p|^4 
+ 2 p^2 \widehat{V} (p)} \, a_p^* a_p \,.\end{split}  \]
This shows that the wave function $U_N^* T_\tau \Omega \in L^2_s (\L^N)$ is a good approximation for the ground state of $H^{\beta=0}_N$ and it allows us to read off the ground state energy 
\begin{equation}\label{eq:ENmf} E_N^{\beta = 0} \simeq  \frac{(N-1)}{2} \widehat{V} (0) - \frac{1}{2} \sum_{p \in \L^*_+} \Big[ p^2 + \widehat{V} (p) -\sqrt{|p|^4 + 2 p^2 \widehat{V} (p)} \Big]\,. \end{equation} 
Moreover, it implies that the low-energy excitation spectrum of $H_N^{\beta = 0} - E_N^{\beta=0}$ consists (up to corrections vanishing as $N \to \infty$) of the eigenvalues 
\begin{equation}\label{eq:excmf} \sum_{p\in \L^*_+} n_p \sqrt{|p|^4 + 2 p^2 \widehat{V} (p)}\,, \end{equation} 
with $n_p \in \bN$, for all $p \in \L^*_+$. Rigorous justifications of (\ref{eq:ENmf}), (\ref{eq:excmf}) for the mean-field regime have been obtained in \cite{S,GS,LNSS,DN} (recently, higher order expansions of the energy have been given in \cite{P3,BPS}). 

For $\beta > 0$, the analysis is more difficult. In fact, for $\beta \geq 1/2$, cubic and quartic terms produce relevant contributions to the energy, and can certainly not be neglected. The main difference with respect to the case $\beta = 0$ is that correlations among particles, which are negligible in the mean-field limit, become here more important. To model correlations, we fix $\ell_0 > 0$ small enough (but independent of $N$) and we consider (similarly to (\ref{eq:neu1})) the ground state solution of the Neumann problem 
\begin{equation}\label{eq:neu2} \Big[ -\Delta + \frac{N^{3\beta}}{2N} V (N^\beta x) \Big] f_{N,\ell_0} = \lambda_{\ell_0} f_{N,\ell_0} \end{equation}
on the ball $|x| \leq \ell_0$, with the normalization $f_{N,\ell_0} (x) = 1$, if $|x| = \ell_0$. We extend $f_{N,\ell_0}$ to $\L$, setting $f_{N,\ell_0} (x) = 1$ for $|x| > \ell_0$ and we define $w_{N,\ell_0} = 1 - f_{N,\ell_0}$. For $p \in \L^*_+$, we consider the coefficients $\eta_p = -N \widehat{w}_{N,\ell_0} (p)$ and we introduce the generalized Bogoliubov transformation 
\begin{equation}\label{eq:Teta} T_\eta = \exp \Big[ \frac{1}{2} \sum_{p \in \L^*_+} \eta_p (b_p^* b_{-p}^* - b_p b_{-p}) \Big]\,. \end{equation} 
To understand (at least heuristically) the choice of the sequence $\eta$, we can write 
\[ T_\eta \Omega \simeq C \exp \Big[ \frac{1}{2} \int dx dy \, \check{\eta} (x - y) a_x^* a_y^* \Big]  \Omega \]
for a constant $C \in \bR$, making sure that normalization is preserved. This leads to 
\begin{equation} \label{eq:eta-dy}
\begin{split}  \big( U_N^* T_\eta \Omega \big) (x_1, \dots , x_N) &\simeq C e^{\frac{1}{N} \sum_{i<j}^N \check{\eta} (x_i - x_j)} \\ &= C \prod_{i<j}^N e^{\frac{1}{N} \check{\eta} (x_i - x_j)} \simeq C \prod_{i<j}^N \big[ 1 + \frac{1}{N} \check{\eta} (x_i - x_j) \big] \end{split} \end{equation} 
and (comparing with a wave function of the form (\ref{eq:psiNL})) explains the choice of $\eta$.

With $T_\eta$, we define the renormalized excitation Hamiltonian 
\[ \cG_{N,\beta} = T_\eta^* \cL_{N,\beta} T_\eta = T_\eta^* U_N H_N^\beta U_N^* T_\eta\,. \]
Similarly to what we did in (\ref{eq:cLN}) for $\cL_{N,\beta}$, we can also decompose the operator $\cG_{N,\beta}$ in constant, quadratic, cubic and quartic terms. Through conjugation with $T_\eta$, we extracted important contributions to the energy from the quartic terms in (\ref{eq:cLN}). As a consequence, the constant term in $\cG_{N,\beta}$ is now closer to the true ground state energy, compared with the constant term in (\ref{eq:cLN}); the vacuum expectations of $\cL_{N,\beta}$ and of $\cG_{N,\beta}$ are given, to leading order, by $N \widehat{V} (0)/2$ and, respectively, $4\pi N \frak{a}^\beta_N$; the difference is of order $N^\beta$, which is exactly the energy carried by the correlations generated by $T_\eta$\footnote{With $\check{\eta} (x) \simeq C \chi (|x| \leq \ell_0)/(|x| + N^{-\beta})$, we expect that $T_\eta$ creates order $\| \eta \|^2_2 \simeq 1$ excitations, with an energy of order $\| \eta \|_{H^1}^2 \simeq N^\beta$.}.

For $\beta \in (0;1)$, it is possible to show that cubic and quartic terms in $\cG_{N,\beta}$ are negligible, in the limit of large $N$. Thus, in this case, the renormalized excitation Hamiltonian $\cG_{N,\beta}$ can be approximated by its quadratic component and can therefore be diagonalized by means of another (generalized) Bogoliubov transformation $T_\tau$, similar to (\ref{eq:Ttau}), correcting the energy at order one. Following this strategy, it was shown in \cite{BBCS3} that, for $\beta \in (0;1)$, the wave function $U_N^* T_\eta T_\tau \Omega \in L^2_s (\L^N)$ is a good approximation for the ground state vector of $H_N^\beta$, that the ground state energy is given by 
\begin{equation}\label{eq:Ebeta} \begin{split} E_N^\beta \simeq \; &4\pi (N-1) \frak{a}_N^\beta -\frac{1}{2}\sum_{p\in\Lambda^*_+} \Big[ p^2+ \widehat{V} (0) - \sqrt{|p|^4 + 2 p^2 \widehat{V} (0)} - \frac{\widehat{V}^2 (0)}{2p^2}\Big] , \end{split} \end{equation}    
and that the low-energy excited eigenvalues of $H_N^\beta - E_N^\beta$ have the form 
\begin{equation*}\label{eq:excbeta} \sum_{p\in\Lambda^*_+} n_p \sqrt{|p|^4+2p^2 \widehat{V} (0)}\,,  \end{equation*}
with $n_p \in \bN$ for all $p \in \L^*_+$, up to errors that vanish as $N \to \infty$. Here, $\frak{a}_N^\beta$ denotes $N$ times the scattering length of the potential $N^{3\beta-1} V (N^\beta \cdot)$ and can be recovered through the finite Born series \footnote{Comparing with (\ref{eq:ENmf}), we conclude that contributions to $\frak{a}_N^\beta$ associated with $k \geq 2$, which are relevant for $\beta \geq 1/2$, emerge from cubic and quartic terms in (\ref{eq:cLN}).} 
\begin{equation} \label{bN}
\begin{split}
8\pi \frak{a}_N^\beta = \; & \widehat{V} (0)-\frac{1}{2N}\sum_{p\in\Lambda^*_+}  \frac{\widehat{V}^2 (p/N^\beta)}{p^2} \\
    &+\sum_{k=2}^{m_\beta}\frac{(-1)^{k}}{(2N)^{k}}\sum_{p\in\Lambda^*_+}\frac{\widehat{V} (p/N^\beta)}{p^2} \\ &\quad \times \sum_{q_1, q_2, \dots, q_{k-1}\in\Lambda^*_+ }\frac{\widehat{V} ((p-q_1)/N^\beta)}{q_1^2}\left(\prod_{i=1}^{k-2}\frac{\widehat{V} ((q_i-q_{i+1})/N^\beta)}{q^2_{i+1}}\right)  \widehat{V} (q_{k-1}/N^\beta)\,,
\end{split}
\end{equation}
where the order $m_\beta$ is chosen so large that the error is much smaller than $1/N$ and therefore only produces negligible contributions when inserted in (\ref{eq:Ebeta}). 

In the Gross-Pitaevskii regime (\ie for $\beta = 1$), even after renormalization with the Bogoliubov transformation $T_\eta$, cubic and quartic terms in $\cG_N^\text{GP} = \cG_{N,\beta=1}$ are still important, they cannot be neglected. In this case, following \cite{BBCS3}, we need to perform a second renormalization of the excitation Hamiltonian, this time conjugating it with a unitary operator having the form
\[ S = \exp \Big[ \frac{1}{\sqrt{N}} \sum_{r \in P_H, v \in P_L} \eta_r \big(\sinh (\eta_v) b_{r+v}^* b_{-r}^* b_{-v}^* + \cosh (\eta_v) b_{r+v}^* b_{-r}^* b_v - \text{h.c.} \big) \Big] \]
given by the exponential of a cubic expression in (modified) creation and annihilation operators (here, $P_H, P_L$ are appropriately defined sets of high and low momenta). 
This leads to the (twice) renormalized excitation Hamiltonian $\cM^\text{GP}_N = S^* T^*_\eta U_N H_N^{\text{GP}} U_N T_\eta S$. Through conjugation with $S$, we extracted additional contributions from the cubic and the quartic terms in $\cG^\text{GP}_{N}$. At this point, the remaining cubic and quartic terms in $\cM^\text{GP}_N$ are small and we can focus on its quadratic part. Diagonalization through a last Bogoliubov transformation $T_\tau$ produces the good ansatz $U_N^* T_\eta S T_\tau \Omega \in L^2_s (\L^N)$ for the ground state vector and leads to the bounds (\ref{eq:ENGP}) for the ground state energy
and (\ref{eq:excGP}) for the excited eigenvalues.

The approach that we discussed in this section and that led to the estimates (\ref{eq:ENGP}),(\ref{eq:excGP}) requires $V \in L^3 (\bR^3)$ (as well as the standard assumptions that $V$ is non-negative, radial and of short range). As recently shown in \cite{NT} (for the case of a Bose gas trapped by an external potential), it could be extended to $V \in L^1 (\bR^3)$ with a slightly different choice of the coefficients $\eta_p$ in (\ref{eq:Teta}) (the stronger condition $V \in L^3 (\bR^3)$ is only important to control properties of the solution of the Neumann problem (\ref{eq:neu2})). On the other hand, this approach cannot be easily extended to particles interacting through a hard-sphere potential. 
The problem is that it is difficult to impose the hard sphere condition on states that are defined through the action of the unitary transformations $T_\eta , T_\tau , S$. Keeping for example $\check{\eta} = -N w_{N,\ell_0}$, as defined after (\ref{eq:neu2}), we would have $\check{\eta} (x)  = -N$ for $|x| < \frak{a}/N$ and, according to the heuristic identity (\ref{eq:eta-dy}), $(U_N^* T_\eta \Omega) (x_1, \dots , x_N)$ would be far from zero, 
even when particles are closer than allowed by the hard-sphere condition (in this case, of course, the last approximation of the exponential in (\ref{eq:eta-dy}) is not valid).  

In the next section, we are going to present an alternative approach that we recently developed to obtain an upper bound for the ground state energy matching (\ref{eq:ENGP}), for Bose gases interacting through hard-sphere potential.

\section{Upper bound for the energy of hard-spheres in the Gross-Pitaevskii limit} 
\label{sec:new}

In this section, we consider a gas of $N$ hard spheres, with radius $\frak{a}_N = \frak{a}/N$, moving on the three dimensional unit torus $\L = [0;1]^3$. We are interested in the ground state energy of the system, which is defined as  
\[ E_N^\text{HS} = \inf  \, \frac{\big\langle \Psi_N , \sum_{j=1}^N -\Delta_{x_j} \Psi_N \big\rangle}{\| \Psi_N \|^2}  \]
with the infimum taken over $\Psi_N \in L^2_s (\L^N)$, $\Psi_N \not = 0$, satisfying the hard-sphere condition 
\begin{equation}\label{eq:hs-con} 
\Psi_N (x_1, \dots, x_N) = 0, \text{ if there exist } i,j \in \{ 1, 2, \dots , N \} \text{ with } i \not = j  \text{ and } |x_i -x_j| \leq \frak{a}/N.
\end{equation} 

\begin{theorem}\label{thm:main} 
There exist $C,\eps > 0$ such that 
\begin{equation}\label{eq:main}  E^\text{HS}_N \leq \; 4 \pi \frak{a} (N  -1) \, +\,  e_\L \aa^2 -\frac 12 \sum_{p \in \L^*_+} \bigg[ p^2 + 8 \pi \aa - \sqrt{|p|^4 + 16 \pi \aa p^2} - \frac{(8 \pi \aa)^2}{2p^2}\bigg] +  C N^{-\eps}
\end{equation} 
with $e_\Lambda$ as defined in (\ref{eq:eLambda0}). 
\end{theorem}

As usual for upper bounds, to prove Theorem \ref{thm:main} we need to find a trial state $\Psi_N \in L^2_s (\L^N)$ satisfying the hard-sphere condition (\ref{eq:hs-con}), whose energy matches the r.h.s. 
of (\ref{eq:main}). 

Summarizing the content of the previous sections, we have discussed two possible approaches to construct an appropriate trial state. Following the Dyson-Jastrow approach presented in Section \ref{sec:dyson}, we could consider a trial function having the form 
\begin{equation}\label{eq:DJ} \Psi_N (x_1, \dots , x_N) = \prod_{i<j}^N f_{N,\ell} (x_i - x_j) \end{equation} 
with $f_{N,\ell}$ describing two-body correlations, up to the distance $1/N \ll \ell \ll 1$. A nice feature of (\ref{eq:DJ}) is the fact that it automatically satisfies the hard-sphere condition (\ref{eq:hs-con}) (assuming of course that $f_{N,\ell} (x) = 0$, for $|x| < \frak{a}/N$). On the other hand, to show a bound of the form (\ref{eq:main}), resolving the energy up to errors that vanish in the limit $N \to \infty$, one would need to choose $\ell$ comparable with the size of the box, of order one. For such values of $\ell$, it seems extremely difficult to control the product (\ref{eq:DJ}) and to compute its energy with sufficient accuracy. Following the Bogoliubov approach, we could instead consider a trial state of the form 
\begin{equation}\label{eq:Bog} \Psi_N = U_N^* T_\eta S T_\tau \Omega \end{equation} 
like those that have been considered in \cite{BBCS3,BBCS4} to prove (\ref{eq:ENGP}) for integrable potentials. Because of the nice algebraic properties of Bogoliubov transformations, these states are much more accessible to computation than (\ref{eq:DJ}). In particular, they allow us to create and to control correlations at all length scales (this is the reason why this approach has been successfully applied in \cite{BCS} and, in a somehow different form, also in \cite{GA,ESY,YY} to compute the upper bound for integrable potentials in the thermodynamic limit). As indicated at the end of Section \ref{sec:bog}, however, it seems very difficult to impose the hard-sphere condition on (\ref{eq:Bog}). 

Our idea to show Theorem \ref{thm:main} is to combine the Dyson-Jastrow and the Bogoliubov  approaches, using a Jastrow factor to capture correlations at short distances and a Bogoliubov transformation to describe them at large distances. In other words, we consider a trial state having the form
\begin{equation}\label{eq:trial} \Psi_N (x_1, \dots , x_N) = \Phi_N (x_1, \dots , x_N) \cdot \prod_{i<j}^N f_{N,\ell} (x_i - x_j) \end{equation} 
where $f_{N,\ell}$ is chosen (similarly to (\ref{eq:neu1}), but now the hard core has the radius $\frak{a}/N$) as the ground state of the Neumann problem 
\begin{equation}\label{eq:neu3}
-\Delta f_{N,\ell} = \lambda_{N,\ell} f_{N,\ell} ,
\end{equation} 
on the ball $|x| \leq \ell$, with $f_{N,\ell} (x) = 0$ for all $|x| \leq \frak{a}/N$. Choosing $\ell \gg N^{-1}$ but small enough, we take care of the hard-sphere conditions and, at the same time, we keep computations involving the Jastrow factor simple. 
%Correlations at length scales larger than $\ell$, on the other hand, will be implemented by the wave %function $\Phi_N$, which will be later defined through appropriate (generalized) Bogoliubov %transformations. 
The wave function $\Phi_N$, on the other hand, will be defined through appropriate (generalized) Bogoliubov transformations; to resolve the ground state energy to the desired precision, it must produce the correct correlations on all length scales larger than $\ell$. Because of the presence of the Jastrow factor, however, it does not need to satisfy hard-sphere conditions. We find it convenient to choose $\Phi_N$ with $\| \Phi_N \| = 1$; as a consequence, $\Psi_N$ will not be normalized in $L^2_s (\L^N)$ (but we will show later that $\| \Psi_N \| \to 1$, as $N \to \infty$). 

Let us now compute the energy of the trial state (\ref{eq:trial}). We find 
%%%%%%
\[ \begin{split} 
\frac{-\Delta_{x_j} \Psi_N (x_1, \dots , x_N) }{\prod_{i<j}^N f_{N,\ell} (x_i - x_j) } = \; &\Big[ -\Delta_{x_j}  - 2 \sum_{i \not = j}^N \frac{\nabla f_{N,\ell} (x_j - x_i)}{f_{N,\ell} (x_j - x_i)} \cdot \nabla_{x_j} \Big] \Phi_N (x_1, \dots , x_N) \\ &+  \sum_{i \not = j}^N \frac{-\Delta f_{N,\ell} (x_j - x_i)}{f_{N,\ell} (x_j - x_i)}  \Phi_N (x_1, \dots , x_N) \\ &- \sum^N_{i,m,j} \frac{\nabla f_{N,\ell} (x_j - x_i)}{f_{N,\ell} (x_j. -x_i)} \cdot \frac{\nabla f_{N,\ell} (x_j - x_m)}{f_{N,\ell} (x_j - x_m)}  \Phi_N (x_1, \dots , x_N)\,,
\end{split} \]
where the sum in the last term runs over $i,j, m \in \{1, \dots , N \}$ all different (notice that, for $\Phi_N \equiv 1$, the first term vanishes and the other two appeared already in (\ref{eq:dys-comp})).
%%%%%%%%%%%%%%%%%%%%%%
Integrating by parts and using (\ref{eq:neu3}), we conclude that 
%%%%%%%%%%%%%%%
\begin{equation}\label{eq:en-psi} \begin{split}  \langle \Psi_N , &\sum_{j=1}^N -\Delta_{x_j} \Psi_N \rangle \\ = \; &\sum_{j=1}^N \int |\nabla_{x_j} \Phi_N ({\bf x})|^2    \prod_{i<j}^N f_{N,\ell}^2 (x_i - x_j)  d{\bf x} \\ &+ \sum_{i<j}^N 2\lambda_{N,\ell} \int \chi_\ell (x_i - x_j) |\Phi_N ({\bf x})|^2 \prod_{i<j}^N f_{N,\ell}^2 (x_i - x_j) d{\bf x} \\ &- \sum_{i,j,m} \int \frac{\nabla f_{N,\ell} (x_j - x_i)}{f_{N,\ell} (x_j - x_i)} \cdot \frac{\nabla f_{N,\ell} (x_j - x_m)}{f_{N,\ell} (x_j - x_m)} |\Phi_N ({\bf x})|^2 \prod_{i<j}^N f^2_{N,\ell} (x_i - x_j) d{\bf x} \,,
\end{split} \end{equation} 
where ${\bf x} = (x_1, \dots ,x_N) \in \L^N$. 
%%%%%%%%%%%%%%%%%%%%
The three-body term on the last line turns out to be negligible, with the appropriate choice of $\ell$ and of the wave function $\Phi_N$. In fact, using permutation invariance and the operator inequality 
\[ \pm W (x_1 - x_2) W (x_1 - x_3) \leq C \| W \|_r^2 (1- \Delta_{x_1})(1-\Delta_{x_2}) (1-\Delta_{x_3}) \]
valid for all $r > 1$, we can bound its contribution by
\begin{equation}\label{eq:num-bd} \begin{split} \Big| \frac{1}{\| \Psi_N \|^2} \sum_{i,j,m} \int &\frac{\nabla f_{N,\ell} (x_j - x_i)}{f_{N,\ell} (x_j - x_i)} \cdot \frac{\nabla f_{N,\ell} (x_j - x_m)}{f_{N,\ell} (x_j - x_m)} |\Phi_N ({\bf x})|^2  \prod_{i<j}^N f_{N,\ell}^2 (x_i - x_j) d{\bf x} \Big| \\ &\hspace{1cm} \leq \frac{C N^3}{\| \Psi_N \|^2}  \int |\nabla f_{N,\ell} (x_1 - x_2)||\nabla f_{N,\ell} (x_1 - x_3)| |\Phi_N  ({\bf x})|^2 d{\bf x} \\ &\hspace{1cm} \leq \frac{C N^3}{\| \Psi_N \|^2} \| \nabla f_{N,\ell} \|_r^2 \,  \langle \Phi_N , (1- \Delta_{x_1}) (1-\Delta_{x_2}) (1-\Delta_{x_3}) \Phi_N \rangle \,.\end{split} \end{equation} 
For a rough estimate, we can approximate $f_{N,\ell} \simeq 1 - \frak{a} \chi_\ell (x) / (N|x|)$ 
(which is what we would find cutting off the solution of the zero energy scattering equation at $|x| = \ell$). This gives \begin{equation} \label{eq:nablaw} \| \nabla f_{N,\ell} \|^2_r \leq C \ell^{6-4r} / N^2, \end{equation} for $1 \leq r < 3/2$. 

Although we still need to define $\Phi_N$, we already know that it will have to describe two-body correlations on scales $|x| \geq \ell$. To get an idea of the size of the expectation on the r.h.s. of (\ref{eq:num-bd}), we can therefore replace $\Phi_N$ (after integrating out the other $(N-3)$ variables) by the three-body wave function 
\begin{equation}\label{eq:apphi} \Big[ 1 - \frac{\frak{a}}{N (|x_1- x_2| + \ell)} \Big]   \Big[ 1 - \frac{\frak{a}}{N (|x_2- x_3| + \ell)} \Big]   \Big[ 1 - \frac{\frak{a}}{N (|x_1- x_3| + \ell)} \Big] \,. \end{equation}  
This leads to the bounds
\begin{equation}\label{eq:derPhi} \begin{split} 
\langle \Phi_N, (-\Delta_{x_1}) \Phi_N \rangle &\lesssim \frac{1}{N\ell} \\
\langle \Phi_N, (-\Delta_{x_1}) (-\Delta_{x_2}) \Phi_N \rangle &\lesssim \frac{1}{N^2 \ell^3} \\
\langle \Phi_N, (-\Delta_{x_1}) (-\Delta_{x_2}) (-\Delta_{x_3}) \Phi_N \rangle &\lesssim \frac{1}{N^3 \ell^4} \,.
\end{split} \end{equation} 
Since $\Phi_N$ describes two-body correlations, it is not surprising that adding a second Laplacian costs more in the estimates than introducing the first and the third. Of course, so far (\ref{eq:derPhi}) are only heuristic bounds; to make our arguments rigorous, after proper choice of $\Phi_N$ we should verify that (\ref{eq:derPhi}) really holds true (the rigorous bounds are proven in \cite{BCOPS}).

As for the norm of $\Psi_N$, appearing in the denominator in (\ref{eq:num-bd}), we set $u_{N,\ell} = 1 - f_{N,\ell}^2$ and we bound 
\begin{equation}\label{eq:prod-bd} \prod_{i<j}^N f_{N,\ell}^2 (x_i - x_j) \geq 1 - \sum_{i<j}^N u_{N,\ell} (x_i - x_j)\,. \end{equation} 
This implies that 
\begin{equation}\label{eq:Psinorm}  \| \Psi_N \|^2 \geq 1 - \frac{N(N-1)}{2} \int u_{N,\ell} (x_1 - x_2) |\Phi_N ({\bf x} ) |^2 d{\bf x}\,. \end{equation}
With the heuristic approximation $f_{N,\ell} \simeq 1- \frak{a} \chi_\ell (x)/(N|x|)$, we find $\| u_{N,\ell} \|_1 \leq C \ell^2 /N$. From the operator inequality 
\[ \pm W (x_1 - x_2) \leq C \| W \|_1 (1-\Delta_{x_1})^{3/4+\delta/2} (1- \Delta_{x_2})^{3/4+\delta/2} \]
valid for any $\delta > 0$, and using again (\ref{eq:apphi}) to heuristically estimate \[ \langle \Phi_N, (-\Delta_{x_1})^{3/4+\delta/2} (-\Delta_{x_2})^{3/4+\delta/2} \Phi_N \rangle \lesssim  \frac{1}{N^2 \ell^{2+\delta}} ,\] we conclude that 
\begin{equation}\label{eq:Psinorm2} \begin{split} 
\Big|  \frac{N(N-1)}{2} \int u_{N,\ell} &(x_1 - x_2) |\Phi_N ({\bf x} ) |^2 d{\bf x} \Big| \\ &\leq C N^2 \| u_{N,\ell} \|_1 \, \langle \Phi_N, (1- \Delta_{x_1})^{3/4+\delta} (1-\Delta_{x_2})^{3/4+\delta} \Phi_N \rangle  \\ &\leq C N \ell^2 \Big[ 1 + \frac{1}{N^2 \ell^{2 + 2\delta}} \Big]\,, \end{split}\end{equation}
which in particular implies (choosing $\delta > 0$ small enough) that $\| \Psi_N \| \geq 1/2$. 

From (\ref{eq:num-bd}), we expect therefore that for every $\delta > 0$ there exists $C > 0$ with 
\[ \begin{split} \Big| \frac{1}{\| \Psi_N \|^2} \sum_{i,j,m} \int \frac{\nabla f_{N,\ell} (x_j - x_i)}{f_{N,\ell} (x_j - x_i)} &\cdot \frac{\nabla f_{N,\ell} (x_j - x_m)}{f_{N,\ell} (x_j - x_m)} |\Phi_N ({\bf x})|^2  \prod_{i<j}^N f_{N,\ell}^2 (x_i - x_j) d{\bf x} \Big| \\ &\hspace{1.5cm} \leq C N \ell^{2-\delta} \Big[ 1 + \frac{1}{N^2 \ell^3} \Big] \leq C N\ell^{2-\delta} + \frac{C}{N\ell^{1+\delta}}\,.  \end{split} \] 
Fixing $\delta > 0$ small enough, we conclude that this term is negligible, in the limit $N \to \infty$, for every choice of $\ell$ satisfying $N^{-1} \ll \ell \ll N^{-1/2}$.

So, let us focus on the first two terms on the r.h.s. of (\ref{eq:en-psi}). We define 
\begin{equation}\label{eq:kinpot} \begin{split} E_\text{kin} (\Phi_N) &= \sum_{j=1}^N \int |\nabla_{x_j} \Phi_N ({\bf x})|^2 \prod_{i<j}^N f_{N,\ell}^2 (x_ i -x_j) d{\bf x}  \\
E_\text{pot} (\Phi_N) &= \sum_{i<j} 2\lambda_{N,\ell} \int \chi_\ell (x_i - x_j) |\Phi_N ({\bf  x})|^2 \prod_{i<j}^N f_{N,\ell}^2 (x_ i -x_j) d{\bf x}\,. \end{split}  \end{equation} 
Complementing (\ref{eq:prod-bd}) with the upper bound 
\[ \prod_{i<j}^N f_{N,\ell}^2 (x_i - x_j) \leq 1 - \sum_{i<j} u_{N,\ell} (x_i - x_j) + \frac{1}{2} \sum_{\substack{i<j , m < n: \\ (i,j) \not = (m,n)}} u_{N,\ell} (x_i - x_j) u_{N,\ell} (x_m - x_n) \]
the kinetic term defined in (\ref{eq:kinpot}) can be estimated by 
\begin{equation} \label{eq:Ekin1} \begin{split} E_\text{kin} &(\Phi_N) \\ \leq \; &N \int |\nabla_{x_1} \Phi_N (\textbf{x})|^2 d{\bf x} - N \int |\nabla_{x_1} \Phi_N ({\bf x})|^2 \sum_{i<j} u_{N,\ell} (x_i - x_j)  \, d {\bf x}  \\ &+ \frac{N}{2} \int |\nabla_{x_1} \Phi_N ({\bf x})|^2 \sum_{\substack{i<j, m<n :\\ (i,j) \not = (m,n)}} u_{N,\ell} (x_i - x_j) u_{N,\ell} (x_m - x_n) \, d{\bf x}  \\ = \; &N \int |\nabla_{x_1} \Phi_N ({\bf x}) |^2 (1- (N-1) u_{N,\ell} (x_1 - x_2)) d{\bf x} \\ &- \frac{N (N-1)(N-2)}{2} \int |\nabla_{x_1} \Phi_N ({\bf x})|^2 (1 - (N-3) u_{N,\ell} (x_1 - x_2)) u_{N,\ell} (x_3 - x_4) \, d{\bf x} \\ &+ \cE_\text{kin} \end{split} \end{equation} 
with a small error $\cE_\text{kin}$ which can be estimated arguing similarly to what we did to bound the r.h.s. of (\ref{eq:num-bd}) and vanishes, as $N \to \infty$, if $N^{-1} \ll \ell \ll N^{-2/3}$. Instead of (\ref{eq:nablaw}), we use here the fact (which can be justified by $f_{N,\ell} \simeq 1 - \frak{a} \chi_\ell (x)/(N|x|)$) that, for any $1 \leq r < 3$,  \[ \| u_{N,\ell} \|_r  \leq C\ell^{3/r-1} /N \,. \]
Moreover, to control $\Phi_N$ we need estimates similar to (\ref{eq:derPhi}), but with derivatives hitting more particles (up to five, in fact); see \cite[Section 4]{BCOPS} for more details.

As for the potential term in (\ref{eq:kinpot}), we obtain 
\begin{equation} \label{eq:Epot1} \begin{split} &E_\text{pot} (\Phi_N) \\ &\leq N (N-1) \lambda_{N,\ell} \int \chi_\ell (x_1 - x_2) f_{N,\ell}^2 (x_1 - x_2) |\Phi_N ({\bf x})|^2 d{\bf x} \\ &\hspace{.4cm}-  \frac{N (N-1)(N-2)(N-3)}{2} \lambda_{N,\ell} \int \chi_\ell (x_1 - x_2) f_{N,\ell}^2 (x_1- x_2) |\Phi_N ({\bf x})|^2 u_{N,\ell} (x_3 - x_4) d {\bf x} \\ &\hspace{.4cm}+ \cE_\text{pot} \end{split} \end{equation} 
for another error $\cE_\text{pot}$, vanishing as $N \to \infty$ (for $N^{-1} \ll \ell  \ll N^{-3/4}$ but small enough); again, details can be found in \cite[Section 4]{BCOPS}. 

Let us introduce the effective $N$-particle Hamilton operator 
\begin{equation}\label{eq:Heff}
H_N^\text{eff} = \sum_{j=1}^N - \Delta_{x_j} + 2 \sum_{i<j}^N \nabla_{x_j} \cdot u_{N,\ell} (x_i - x_j) \nabla_{x_j} + 2\lambda_{N,\ell}  \sum_{i<j}^N \chi_\ell (x_i - x_j) f_{N,\ell}^2 (x_i - x_j) \,.
\end{equation} 
Then, combining (\ref{eq:Ekin1}) and (\ref{eq:Epot1}) with (\ref{eq:Psinorm}), we conclude from (\ref{eq:en-psi}) that 
\[  \begin{split} &\frac{\big\langle \Psi_N, \sum_{j=1}^N -\Delta_{x_j} \Psi_N \big\rangle}{\| \Psi_N \|^2} 
 \\ &\hspace{.3cm} \leq \Big[  1 + \frac{N(N-1)}{2} \int u_{N,\ell} (x_1 - x_2) |\Phi_N ({\bf x})|^2 d{\bf x}  \Big] \\ &\hspace{2cm} \times  \Big[ \langle \Phi_N, H_N^\text{eff} \Phi_N \rangle - \frac{N(N-1)}{2} \langle \Phi_N , \big[ H_{N-2}^\text{eff} \otimes u_{N,\ell} (x_{N-1} - x_N) \big]  \Phi_N \rangle  \Big] \\&\hspace{.7cm} + \cE \end{split} \]
for an error $\cE$, vanishing as $N \to \infty$, if $N^{-1} \ll \ell \ll N^{-3/4}$. Here we brought the second term on the r.h.s. of (\ref{eq:Psinorm}) to the numerator, using (\ref{eq:Psinorm2}) to show that its square is negligible, even after multiplication with quantities of order $N$. We obtain 
\begin{equation}\label{eq:first-part} \begin{split} &\frac{\big\langle \Psi_N, \sum_{j=1}^N -\Delta_{x_j} \Psi_N \big\rangle}{\| \Psi_N \|^2} 
 \\ &\hspace{.8cm} \leq \langle \Phi_N , H_N^\text{eff} \Phi_N \rangle \\ &\hspace{1.1cm} -  \frac{N(N-1)}{2} \Big\langle \Phi_N, \Big\{ \big[ H_{N-2}^\text{eff} - \langle \Phi_N , H_N^\text{eff} \Phi_N \rangle \big] \otimes 
 u_{N,\ell} (x_{N-1} - x_N) \Big\} \Phi_N \Big\rangle + \cE\,.  \end{split} \end{equation}
Eq. (\ref{eq:first-part}) dictates the choice of the wave function $\Phi_N$. To get the best possible upper bound, we should take $\Phi_N$ so that the expectation of $H_N^\text{eff}$ is as small as possible, making sure that the regularity bounds in (\ref{eq:derPhi}) (and the additional bounds needed to control the error terms in (\ref{eq:Ekin1}) and (\ref{eq:Epot1})) are satisfied. 

Since $u_{N,\ell}$ is small, unless particles are very close, we can think of $H_N^\text{eff}$ as a perturbation of the many-body Hamiltonian 
\begin{equation}\label{eq:wtHeff} \wt{H}_N^\text{eff} = \sum_{j=1}^N -\Delta_{x_j} + 2\lambda_{N,\ell} \sum_{i<j}^N \chi_\ell (x_ i - x_j)\,. \end{equation} 
Recalling, from (\ref{eq:lambda}) (but now with $\frak{a}$ replaced by $\frak{a}/N$), that $\lambda_{N,\ell} \simeq 3 \frak{a}/ N\ell^3$, we can consider (\ref{eq:wtHeff}), in good approximation, as an Hamiltonian of the form (\ref{eq:HNbeta}), with $\beta \in (0;1)$ chosen so that $\ell = N^{-\beta}$. As discussed in Section \ref{sec:bog}, we know how to approximate the ground state energy and the ground state wave function of Hamilton operators of this form; we need to introduce Bogoliubov transformations $T_\eta, T_\tau$ as defined in (\ref{eq:Teta}), (\ref{eq:Ttau}) and we have to consider states of the form $U_N^* T_\eta T_\tau \Omega \in L^2_s (\L^N)$. This leads to the expression (\ref{eq:Ebeta}) for the ground state energy. 

Unfortunately, considering the Hamilton operator (\ref{eq:wtHeff}) is not enough, the difference to (\ref{eq:Heff}) is not small and needs to be taken into account. Still, we can apply the rigorous version of Bogoliubov theory that has been developed in \cite{BBCS2} to determine the spectrum of (\ref{eq:wtHeff}) also to study the ground state energy of (\ref{eq:Heff}) and to construct an approximation for its ground state vector. 

Since $H^\text{eff}_N$ corresponds in (\ref{eq:HNbeta}) to an intermediate regime in (\ref{eq:HNbeta}), with $0< \beta < 1$, the renormalization of the excitation Hamiltonian only involves generalized Bogoliubov transformations, no cubic renormalization is required. The presence of the second term on the r.h.s. of (\ref{eq:Heff}), however, affects the choice of the sequence 
$\eta$, needed in (\ref{eq:Teta}) to define $T_\eta$. Let $\ell_0 > 0$ be sufficiently small but fixed, of order one. It turns out that, for every momentum $p \in \L^*_+$, one can take $\eta_p$ as the Fourier coefficient of the function $\check{\eta} = - N (1 - g_{N,\ell_0})$, with $g_{N,\ell_0} = f_{N,\ell_0} / f_{N,\ell}$, given by the ratio of the two solutions of  (\ref{eq:neu3}), defined on balls of radii $\ell_0$ and $\ell$ (recall that $N^{-1} \ll \ell \ll N^{-3/4}$, while $\ell_0$ is small but fixed, of order one). It is then easy to verify that $g_{N,\ell_0}$ satisfies the partial differential equation 
\begin{equation}\label{eq:scatt2} \Big[ -\Delta - 2 \frac{\nabla f_{N,\ell}}{f_{N,\ell}} \cdot \nabla \Big] g_{N,\ell_0} + \lambda_{N,\ell} \chi_\ell f_{N,\ell}^2 g_{N,\ell_0} = \lambda_{N,\ell_0} \chi_{\ell_0} g_{N,\ell_0}\, . \end{equation} 

To understand this choice of $\eta$, recall from (\ref{eq:eta-dy}) that, at least on the heuristic level, $U_N^* T_\eta \Omega$ is an approximation for the product \[ \prod_{i<j}^N \big[1+\frac{1}{N} \check{\eta} (x_i - x_j) \big] = \prod_{i<j}^N g_{N,\ell_0} (x_i - x_j) = \prod_{i<j}^N \frac{f_{N,\ell_0} (x_i - x_j)}{f_{N,\ell} (x_i - x_j)} . \]
This is exactly what is needed, in (\ref{eq:trial}), to replace $f_{N,\ell}$ by $f_{N,\ell_0}$. This procedure introduces, in our trial state, the missing two-body correlations, up to the scale $\ell_0$. 

A part from this heuristic explanation, the choice of the sequence $\eta$ is determined by the computation of the renormalized excitation Hamiltonian $\cG^\text{eff}_{N} = T_\eta^* U_N H_N^\text{eff} U_N^* T_\eta$, acting on the truncated Fock space $\cF_+^{\leq N}$. Compared with the analysis in \cite{BBCS2} (where the initial Hamilton operator has essentially the form (\ref{eq:wtHeff})), the second term on the r.h.s. of (\ref{eq:Heff}) 
and the presence of the factor $f_{N,\ell}^2$ in the third term on the r.h.s. of (\ref{eq:Heff}) produce new large contributions to $\cG_N^\text{eff}$. The condition that these terms cancel (when combined with the large contributions arising from the conjugation of (\ref{eq:wtHeff})), so that $\cG_N^\text{eff}$ can be well approximated by a quadratic operator, fixes the correct form of $\eta$ (the condition appears essentially as the equation   (\ref{eq:scatt2})).

At last, we need to diagonalize the quadratic part of $\cG_N^\text{eff}$. As explained in Section \ref{sec:bog}, this can be achieved through conjugation with a second generalized Bogoliubov transformation $T_\tau$. This leads us to the trial state $\Phi_N = U_N^* T_\eta T_\tau \Omega \in L^2_s (\L^N)$ for the ground state wave function of the effective Hamilton operator (\ref{eq:Heff}). While $T_\eta$ takes care of correlations on length scales between $\ell$ and $\ell_0$, the final conjugation with $T_\tau$ introduces the last missing two-body correlations, on scales larger than $\ell_0$.

%(there are three different length scales that are relevant in the construction of our trial %state: 1) the solution $f_{N,\ell}$ of (\ref{eq:neu3}), appearing in the product (\ref{eq:trial}), %varies on the scale $N^{-1}$, 2) the function $\check{\eta}$, defined through the solution %of (\ref{eq:scatt2}) varies on the scale $\ell$, 3) the function $\check{\tau}$, defined %through its Fourier coefficients, so that $T_\tau$ diagonalized the quadratic part of $%\cG_N^\text{eff}$ varies on the scale $\ell_0$)

The details of this part of the analysis can be found in \cite[Sections 5 and 6]{BCOPS}. At the end, with this definition of $\Phi_N$, we obtain, on the one hand, the estimate  
\begin{equation}\label{eq:final} \langle \Phi_N , H_N^\text{eff} \Phi_N \rangle = 4 \pi \frak{a} (N  -1) \, +\,  e_\L \aa^2 -\frac 12 \sum_{p \in \L^*_+} \bigg[ p^2 + 8 \pi \aa - \sqrt{|p|^4 + 16 \pi \aa p^2} - \frac{(8 \pi \aa)^2}{2p^2}\bigg] +  C N^{-\eps}\,. \end{equation} 
On the other hand, we can show that $\Phi_N$ satisfies the regularity bounds (\ref{eq:derPhi}) (and also the more involved bounds needed to control error terms 
arising from (\ref{eq:Ekin1}) and (\ref{eq:Epot1})); see \cite[Section 7]{BCOPS}.

Inserting (\ref{eq:final}) on the r.h.s. of (\ref{eq:first-part}), the proof of Theorem \ref{thm:main} is almost complete. What is still missing is a bound showing that the second contribution on the r.h.s. of (\ref{eq:first-part}) is negligible, in the limit $N \to \infty$. This requires some additional work, because the second term on the r.h.s. (\ref{eq:Heff}) affects the coercivity of the excitation Hamiltonian $U_N H_{N-2}^\text{eff} U_N^*$. We skip here further details, which can be found in \cite[Section 8]{BCOPS}.  

\medskip

{\it Acknowledgements.} A.O., G.P. and B.S. gratefully acknowledge support from the European Research Council through the ERC Advanced Grant CLaQS. Additionally, B. S. acknowledges partial support from the NCCR SwissMAP and from the Swiss National Science Foundation through the Grant ``Dynamical and energetic properties of Bose-Einstein condensates''.  G.B., S.C., and A.O. warmly acknowledge  the GNFM Gruppo Nazionale per la Fisica Matematica - INDAM.

\end{document}